\documentclass[12pt]{article}

\usepackage{psfig}

\begin{document}
\title{SURVEY OF ROTATION CURVES FOR NORTHERN SPIRAL EDGE-ON GALAXIES}
\author{Makarov D.I., Karachentsev I.D., Burenkov A.N}
\date{}
\maketitle

\begin{abstract}
Rotational curves are measured for almost complete sample of 308 northern
spiral galaxies using the 6-m telescope. For observations we selected
edge-on galaxies from RFGC catalog with apparent axial ratio $a/b > 8$,
angular diameter $a < 2'$, and declinations $\delta > +38^\circ$ above the
Arecibo belt. About 97\% of the galaxies are detected in the H-alpha line.
For them a typical rotation curve extends till 87\% of the standard
galaxy radius.
The rotation curve amplitude shows close correlation with a width of
the HI line. The median radial velocity for the galaxies is
7800~km/sec, and the median maximum of the rotation curves is 137~km/s.
These new observational data will be used to study cosmic streamings of
galaxies seen on a scale of 200~Mpc.
\end{abstract}

The rotation curves of the galaxies from the
flat galaxy catalog (FGC) (Karachentsev et al. 1993)
have been studied at the Special Astrophysical
Observatory since 1995. The goal of this is to estimate
the rotation amplitudes and redshifts for northern-sky
objects and to analyze the pecular velocity field of
the FGC galaxies. The observing program has
been finished in 1999 (Makarov et al. 1997a, 1997b, 1999, 2000),
which allows us to summarize preliminary results.

The FGC catalog has a high degree
of homogeneity and completeness, which make it one
of the best existing samples for analyzing large-scale
motions.
For the observations, we select northern-sky galaxies with
$\delta>+38^\circ$ with apparent axial ratios $a/b>8$ and
major axes in the range $0.\!'6<a<2'$.
The observations are being carried out with the 6-m
telescope at the Special Astrophysical Observatory
by using an optically efficient prime-focus spectrograph.

These observations
were used to estimate the amplitude and direction of the
bulk motion of FGC galaxies with respect to the cosmic
microwave background (Karachentsev et al. 1995, 1999).
It follows from these studies that
bulk motions of FGC galaxies with an amplitude of
$300\pm80$~km/s are observed on scales of the order of 100~Mpc
toward the concentration of rich Shapley clusters.

\begin{figure}
\centerline{  \psfig{figure=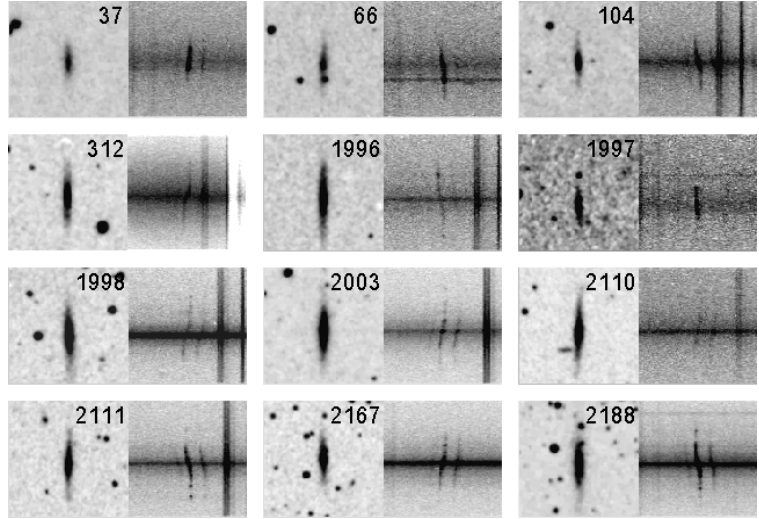,width=10cm}  }
\vspace{1cm}
\par
\centerline{  \psfig{figure=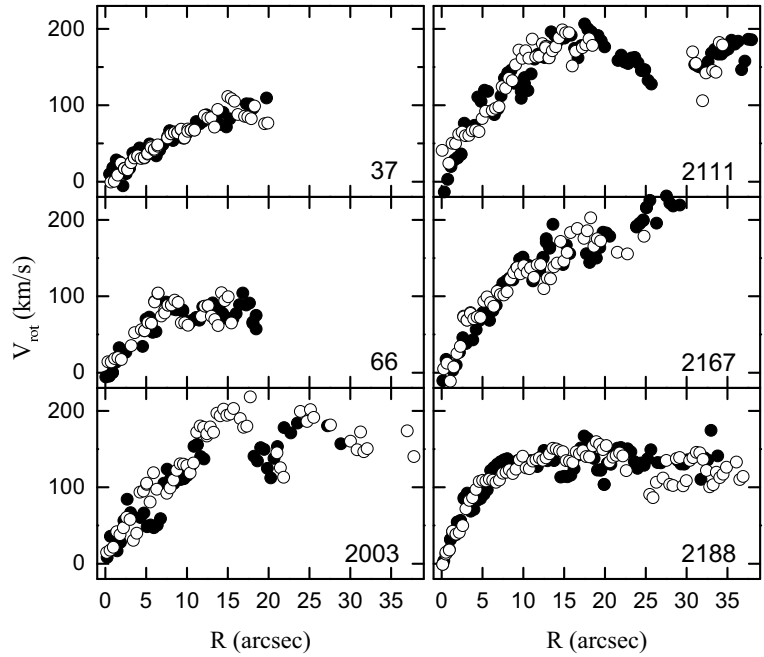,width=10cm}  }
\caption{
Example of POSS image, spectrum near $H_\alpha$ and
rotation curve for some of the investigated galaxies.}
\end{figure}

\begin{figure}
\centerline{
\begin{tabular}{p{6.5cm}p{6.5cm}}
\centerline{\psfig{figure=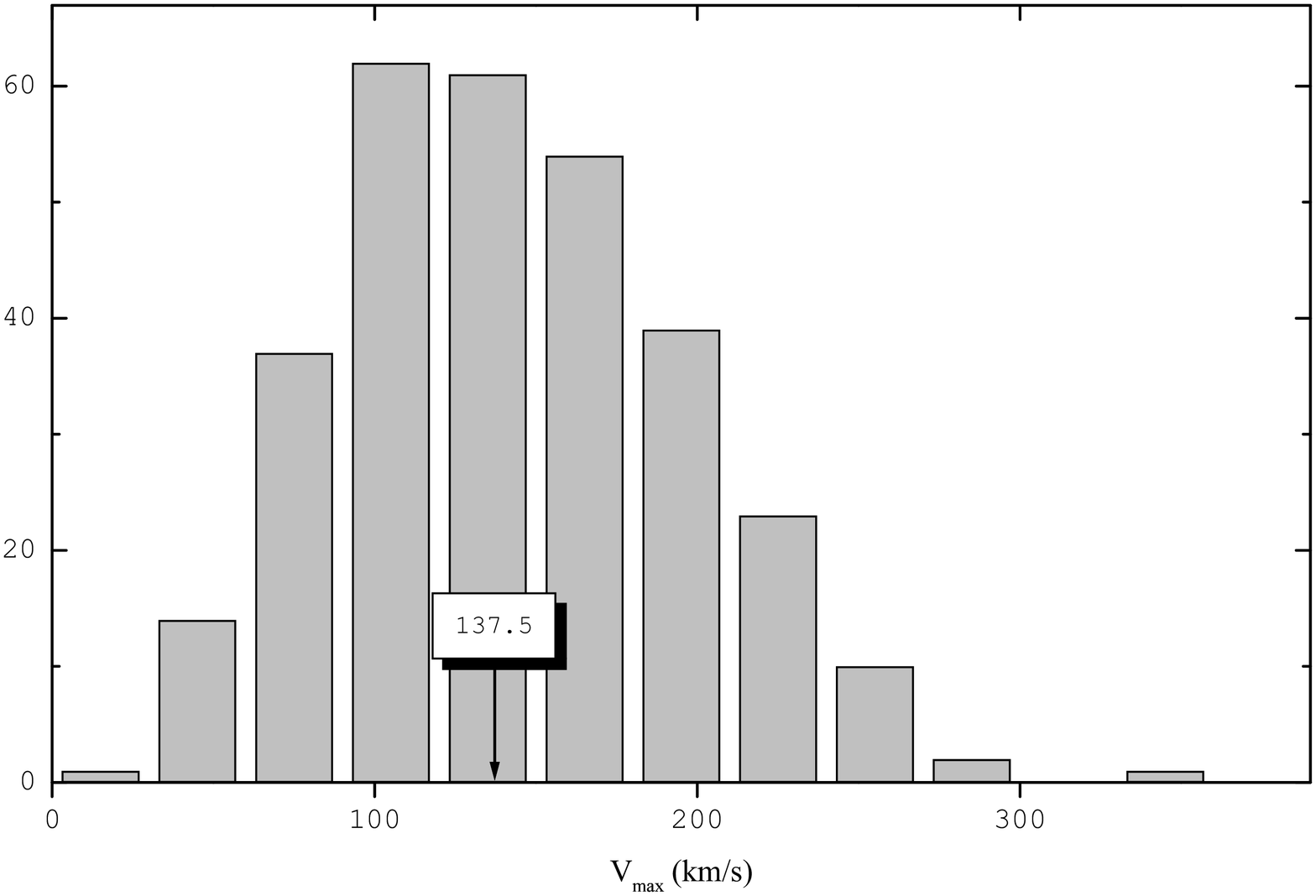,width=6cm}} &
\centerline{\psfig{figure=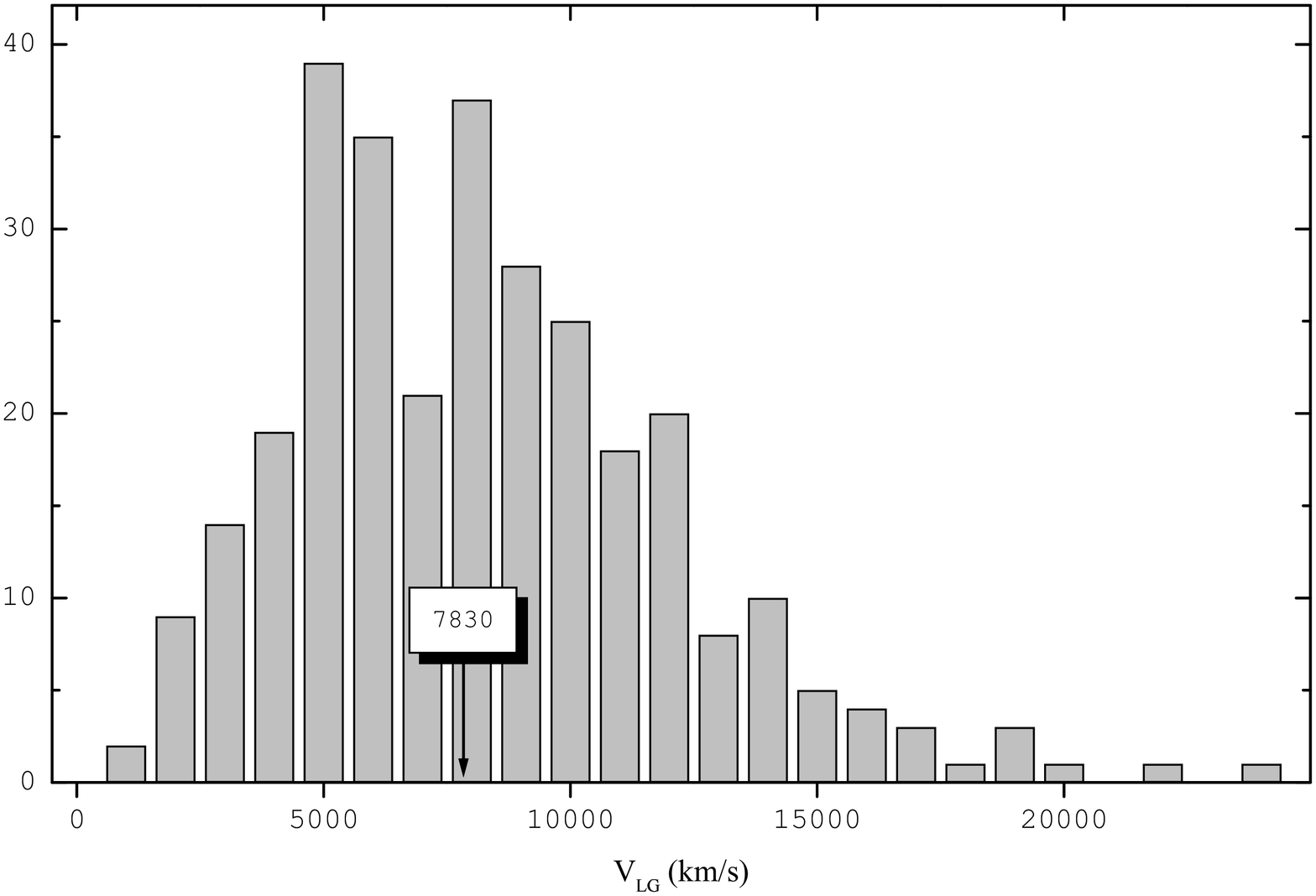,width=6cm}} \\
\caption{The histogram of rotation curve amplitudes.
  The maximum of rotation curve corrected for relativistic widening.
  The median value $V_{\max}=137.5$~(km/s)
  is indicated by the arrow.}  &
\caption{
Histogram of radial velocity corrected for
   Local Group motion.
   The arrow indicates the median value $V_{\mbox{LG}}=7830$~(km/s).}
\end{tabular}
}
\end{figure}

A preliminary analysis for
308 galaxies leads us to the following conclusions:

\begin{itemize}
\item The observation of thin edge-on galaxies are highly efficient.
For more than 97\% of the FGC objects,
the contrast of the emission line is high enough to measure the
typical galaxy rotation amplitude with an error of ~10\%,
which as acceptable in the study of large-scale motions.

\item The extent of the rotation curve for different
galaxies lies in the range 0.45--1.45 with median value equal to
0.87 of the standard radius of a flat galaxy.

\item Several rotation curve of edge-on galaxies appear
signs of large extinction in the plane of the galaxy spiral arms,
which is intersected by the line of sight at small angles.

\item Nevertheless, the galaxies under study show a close correlation
between the rotation amplitudes, as inferred from the $H_\alpha$ and
21-cm HI lines with the rms deviation $\sigma(V_{\rm max})=12$~km/s, which
makes the use of flat galaxies to determine the distances from
Tully-Fisher's diagram (Makarov et al. 1997a) very promising.

\item The median radial velocity corrected to the
Local Group motion equal to 7800 km/s.
The galaxies studied are generally massive spirals with a
median maximum of the rotation curve equal to 137 km/s.
\end{itemize}

These new
observational data will be used to study cosmic streamings
of galaxies seen on a scale of 200~Mpc.

\section*{ACKNOWLEDGMENTS}

We wish to thank V.~V.~Vlasyuk for the help in the observations
and N.~V.~Tyurina and G.~G.~Korotkova for the help in the
data reduction.


\thebibliography{}
\bibitem{} Karachentsev I.D., Karachentseva V.E. and Parnovsky S.L, Astron. Nachr., 1993, 313, 97
\bibitem{} Karachentsev I.D., Karachentseva V.E., Kudrya Yu.N. and  Parnovsky S.L, Astron. Nachr., 1995,316, 369
\bibitem{} Karachentsev I.D., Karachentseva V.E., Kudrya Yu.N. and  Parnovsky S.L, Astron. Zh., 1999, (in press)
\bibitem{} Makarov D.I, Karachentsev I.D., Tyurina N.V. and Kaisin S.S.,  Pis'ma Astron. Zh., 1997a, 23, 509, Astron.  Lett., 1997, 23, 445
\bibitem{} Makarov D.I, Karachentsev I.D., Burenkov A.N. and Tyurina N.V., Pis'ma Astron. Zh., 1997b, 23, 736, Astron. Lett., 1997, 23, 638
\bibitem{} Makarov D.I, Burenkov A.N. and Tyurina N.V., 1999, Pis'ma Astron. Zh., 25, 813, Astron. Lett., 1999, 25, 706
\bibitem{} Makarov D.I, Burenkov A.N. and Tyurina N.V., 2000, (in preparation)
\end{document}